\documentclass[aps,prstab,reprint,groupedaddress]{revtex4-1}
\usepackage{graphicx}
\usepackage{dcolumn}
\usepackage{bm}

\begin{document}
\title{Electrostatic cumulation of high-current electron beams for terahertz sources}

\author{S.V.~Anishchenko}
\email{sanishchenko@mail.ru}
\affiliation{Research Institute for Nuclear Problems\\
Bobruiskaya str., 11, 220030, Minsk, Belarus.}%

\author{V.G.~Baryshevsky}
 \email{bar@inp.bsu.by}
 \affiliation{Research Institute for Nuclear Problems\\
Bobruiskaya str., 11, 220030, Minsk, Belarus.}%

\author{A.A.~Gurinovich}
\email{gur@inp.bsu.by}
\affiliation{Research Institute for Nuclear Problems\\
    Bobruiskaya str., 11, 220030, Minsk, Belarus.}%

\begin{abstract}
The electrostatic cumulation of current density in relativistic
vacuum diodes with ring-type cathodes is described theoretically
and confirmed experimentally. The distinctive feature of the
suggested cumulation mechanism is a very low energy spread of
electrons. As a result of electrostatic cumulation, a thin
relativistic electron beam with a current density of 1 kA/mm$^2$
can be formed. This quantity exceeds typical current density
values in high-current Cherenkov sources by an order of magnitude.
Such a beam can be used as an active medium in high-power
terahertz sources.
\end{abstract}


\maketitle

\section{Introduction}
Generation of high-power radiation is one of the main goals of
vacuum terahertz electronics~\cite{Booske2011,Dhillon2017}. The
progress in this field is strongly connected with production in
high-current accelerators~\cite{Booske2011,Dhillon2017,Garate1987}
of relativistic electron beams with ever-increasing density .

A typical high-current accelerator for a THz source consists of an
axially symmetric cathode and anode~\cite{Garate1987}. The cathode
is hemispherical with  curvature radius of several centimeters and
it is spaced  tens of millimeters from the concave anode. The
accelerator produces about 1~kA of total current at $\sim1$~MV
applied voltage. The hole in the anode "cuts" a small central part
of the accelerated beam ($\sim10$~A), which is guided by magnetic
field with the strength $\sim1$~T. Low efficiency of beam usage
limits applications of such an approach.

The second disadvantage of applying traditional high-current
accelerators is high value of an external guiding field (1 T) that
results in large dimensions of the whole system. High magnetic
fields are necessary to prevent beam expansion due to the Coulomb
repulsion. Meanwhile an alternative mechanism precluding beam
expansion is already well-known. This mechanism doesn't require
external guiding fields. It is based on magnetic cumulation
discovered by Bennett~\cite{Bennet1934}. Magnetic cumulation
provides charged-particle beam intensities as high as
10~GW/mm$^2$~\cite{Morrov1971,Bradley1972}, thus enabling the
laboratory investigation of the extreme state of matter. However,
the large particle energy spread does not enable to use
magnetically cumulated electron beams in terahertz radiation
sources.

This paper considers one more alternative for high-current
electron beam cumulation, namely electrostatic
cumulation~\cite{Anishchenko2017,Anishchenko2014}. Occurring in a
relativistic vacuum diode with a ring-type cathode, this mechanism
doesn't suffer from particle energy spread~\cite{Anishchenko2014}.
Our principle task is to provide  theoretical description of
cumulation mechanism and the experimental verification thereof. We
will show that electrostatic cumulation provides current densities
($\sim1$~kA/mm$^2$) sufficient to seed high-power terahertz
radiation sources.

\section{The phenomenon of electrostatic cumulation}
Electrostatic cumulation was first revealed during modeling of
high-current accelerators~\cite{Anishchenko2014}. The qualitative
picture of electrostatic cumulation can be described as follows.
In a relativistic vacuum diode, electron emission is most intense
from the cathode's edges (Fig. 1). Let us consider electrons
emitted from the inner edge. The Coulomb repulsion causes the
charged particles to rush to the region free from the beam. As a
result, the accelerated motion of electrons toward the anode comes
alongside the radial motion to the cathode's  symmetry axis. As a
result, the high-current beam density increases multifold on the
axis as compared to the average current density in the
cathode-anode gap~\cite{Anishchenko2014,Anishchenko2017}.

\begin{figure}[ht]
    \begin{center}
        \resizebox{85mm}{!}{\includegraphics{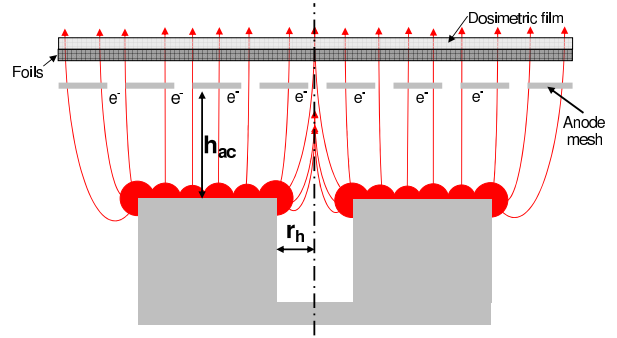}}\\
    \end{center}
    \caption{Cumulation mechanism}\label{Fig.1}
\end{figure}

\begin{figure}[ht]
    \begin{center}
        \resizebox{80mm}{!}{\includegraphics{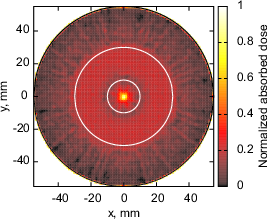}}\\
    \end{center}
    \caption{Electrostatic cumulation: simulation results [cathode radius $r_c=30$mm, anode-cathode gap $h_{ac}=16$mm, hole radius $r_h=10$mm].}\label{Fig.2}
\end{figure}

\begin{figure}[ht]
    \begin{center}
        \resizebox{85mm}{!}{\includegraphics{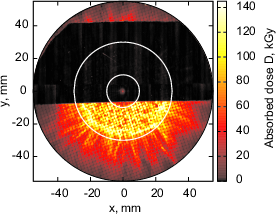}}\\
    \end{center}
    \caption{Electrostatic cumulation: experimental results [cathode radius $r_c=30$mm, anode-cathode gap $h_{ac}=16$mm, hole radius $r_h=10$mm].}\label{Fig.3}
\end{figure}

Figure 2 shows the results of simulations: the dose absorbed by
the anode. The assumed parameters of the cathode are as follows:
cathode radius 30 mm, cathode-anode gap 20 mm, and the radius of
the inner hole 10 mm.  The maximum value of the accelerating
voltage pulse is taken equal to 360 kV and its duration is 130 ns.
The simulated current density in the region of the central spot on
the anode at the moment corresponding to the maximum accelerating
voltage is as large as 10 A/mm2, being 5 times greater than the
average current density of the high-current diode. A typical
radius of the spot is about 1 mm. Thus, the simulation result
indicates the electron-beam cumulation on the axis of a
high-current diode with a ring-type cathode.

The undeniable advantage of electrostatic cumulation is a very low
energy spread of particles due to the laminar flow of charged
particles. In contrast, self-focusing of a beam by its own
magnetic field leads to a turbulent flow, and the charged
particles acquire a significant energy spread. In this case, the
electron flow  is like a compressed relativistic  gas with the
electron temperature $T_e$ determined by the accelerating voltage,
$T_e\sim\frac{q_eU}{k_B}$. (Symbol $k_B$ denotes the Boltzman
constant.)

To obtain information about electron beam parameters, we use a
nanosecond pulse-periodic electron accelerator with a compact
SF6-insulated high-voltage generator  providing pulsed voltage up
to 400 kV in 30 Ohm resistive load with a full width at half
maximum of 130 ns and  rise time of 30 ns~\cite{Anishchenko2017}.
To obtain full-sized imprints of electron beams, we use a
radiochromic dosimetry film placed 3~mm behind the anode mesh made
of stainless steel (the geometrical transparency of the mesh
is~0.77); the cathode-anode gap is 20 mm. The dosimetry film
enables to obtain information about the total absorbed dose over
the beam cross section, caused by passage of charged particles.

Our first experiments had shown that the intense flow of charged
particles on the axis had burned the film through. For this
reason, in further experiments we placed 70~$\mu$m-thick aluminium
foil in front of the dosimetry film to decrease the absorbed dose
(see Fig. 3)~\cite{Anishchenko2017}. This enables us to cut off the  flows of both  the
cathode plasma and the weakly-relativistic electrons produced at
the voltage pulse decay. The experiments conducted with one, two,
and three foils demonstrated that a sharp increase in the absorbed
dose remains in the center. This means that the particle flow
consists of high-energy electrons at the beam axis. In the
experiments with three foil layers cutting off all electrons whose
energy was less than 250 keV, the absorbed dose in the center was
almost four times as large as the average dose across the beam
cross section, showing a good agreement with the simulation
results.

\begin{figure}[ht]
    \begin{center}
        \resizebox{85mm}{!}{\includegraphics{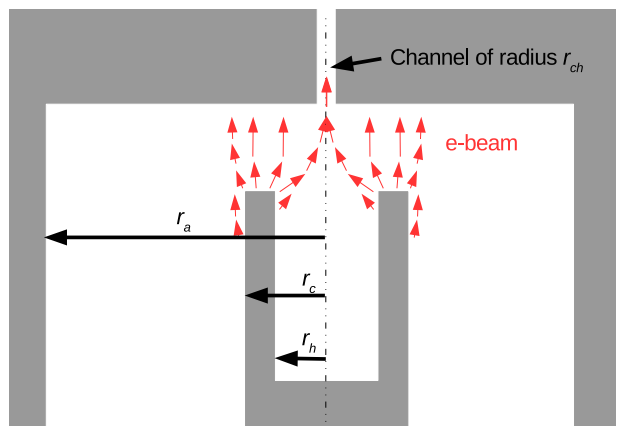}}\\
    \end{center}
    \caption{Extraction of cumulated electron beams.}\label{Fig.4}
\end{figure}

\section{Transport of a dense electron beam at~4~MeV}
Let us note here that both the simulations and the experiments were performed at maximum accelerating voltage 400~kV. The estimates show that it is possible to achieve the beam current density more than 1~kA/mm$^2$ with accelerating voltages of several megavolts. This quantity is several times higher than those obtained in existing high-current terahertz sources producing ~100~kW of radiation power.
The cumulated electron beam can be extracted with the help of a thin channel in the anode (Fig. 4). The beam may be used in terahertz sources.

\begin{figure}[ht]
    \begin{center}
        \resizebox{85mm}{!}{\includegraphics{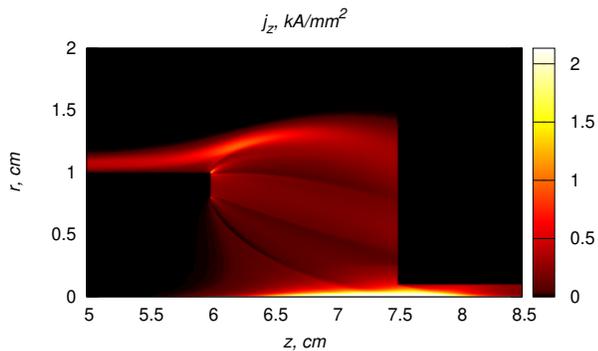}}\\
    \end{center}
    \caption{Sample of simulated current density at 4~MV of applied voltage [channel radius $r_{ch}=1$~mm, channel entrance $z_{ch}=75$~mm].}
    \label{Fig.5}
\end{figure}

Cumulated electron beams are explored with the particle-in-cell
code described in Appendix A. The accelerator geometry consists of
two parts: a cathode and an anode with a channel in its center.
The cathode has an outer radius $r_c$ of 10~mm. The hole radius
$r_h$ has a value 8~mm. The cathode-anode gap $h_{ac}$ is 15~mm.
The channel radius $r_{ch}$ has values 0.5 or 1~mm. The anode
radius $r_a$ is 200~mm.

\begin{figure}[ht]
    \begin{center}
        \resizebox{85mm}{!}{\includegraphics{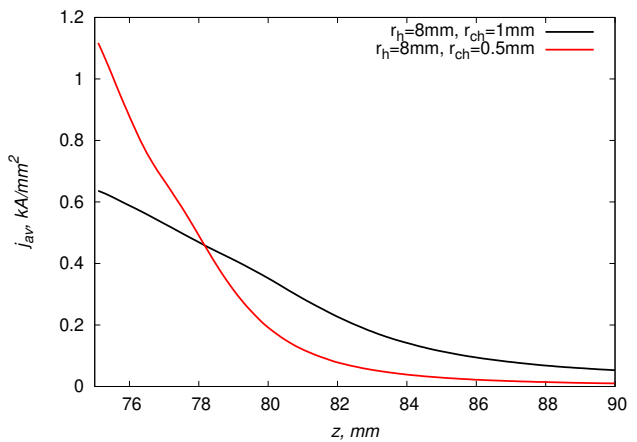}}\\
    \end{center}
    \caption{Average current density as a function of transport distance [channel entrance $z_{ch}=75$~mm].}\label{Fig.6}
\end{figure}

Simulations show that a typical current density in the
channel~$j_{av}$ is about~$1$~kA/mm$^2$ (Fig. 5). The transport
length of cumulated beam lies in the millimeter range (Fig. 6).
Estimations show that the cumulated electron beam, injected in the
slow-wave structure, can lead to the Cherenkov terahertz radiation
of 40---400~MW power if we assume 1---10\% efficiency, which is a
typical value for the high-current Cherenkov generators
\cite{Garate1987,Bugaev1979}.

\section{Conclusions}
In this paper we describe a new cumulation mechanism for
high-current beams in relativistic vacuum diodes with a ring-type
cathode. The basis of the cumulation mechanism is electrostatic
repulsion of electrons emitted from the inner edge of the cathode.
At several hundred kilovolts, the electrostatic cumulation is
verified experimentally. The current density and beam intensity
equal to 1---10~GW/mm$^2$ and 1~kA/mm$^2$, respectively, are
obtained in simulations at~4~MV.

A very low particle energy spread in the region of maximum current
density is a distinctive feature of the described cumulation
mechanism. Due to electrostatic cumulation, a thin relativistic
electron flow with a radial width of 1 mm and beam current density
of 1~kA/mm$^2$ is formed. This quantity is several times higher
than that obtained in existing high-current terahertz sources
producing 0.1---1~MW of radiated power. As a result, electrostatic
cumulation can sufficiently increase the radiation output. One of
the main advantages of high-current terahertz sources with
electrostatic cumulation is the absence of external magnetic
fields causing large dimensions of the whole system.

\appendix
\section{Simulation of high-current electron beams}
Self-consistent simulation of relativistic
motion in high-current diodes is usually
performed using the particle-in-cell method~\cite{Poukey1973,Golovin1989,Roshal1979,Hockney1981,Birdsall1985,Antonsen,Verboncoeur2005}.
 The method is often implemented in a quasi-stationary
approximation \cite{Poukey1973,Golovin1989}. The approximation is
applied when electric and magnetic fields are slowly varying
functions of time. As a result, the displacement currents and
induction fields can be neglected.

As we are going to deal with axially symmetric geometries the
simulation is performed in cylindrical RZ-coordinates. So, let us
introduce the space mesh as follows:

\begin{eqnarray}
& \omega_{RZ}=\{R_i=i\Delta R,R_j=j\Delta Z,\nonumber \\
& i=0,1,...,i_{max},j=0,1,...,j_{max}\} \nonumber 
\end{eqnarray}
and the time mesh
\begin{equation}
\omega_{T}=\{T_n=n\Delta T,n=0,1,...\}.
\end{equation}
Spatial dimensions of the grid cells are set to be equal, i.e.,
$\Delta R=\Delta Z$.

The method implies representation of charged particles' flow as a
collection of macroparticles. Each of them contains a large number
of real charge carriers. Macroparticles have certain spatial
distributions of mass and charge those contribute to the physical
quantities (such as current and charge densities) defined on the
grid.
The latter give us possibility to find electric and magnetic
fields in the grid nodes. Interpolation of the grid fields to
macroparticles' location determines forces acting on charges. As a
result, new particles' locations and velocities can be found. At
this point, all the procedures are repeated again.

Now, let us proceed to the brief description of the algorithm used in our simulations.

Electric field $\vec E$ is expressed by electrostatic potential
$\phi$,
\begin{equation}
\vec E=-\nabla\phi.
\end{equation}
The potential $\phi$ obeys the Poisson equation,
\begin{equation}
\Delta\phi=-4\pi\rho.
\end{equation}
Symbol $\rho$ denotes the charge density.

To solve the Poisson equation,
we use the Jacobi iterative method \cite{Hockney1981},
\small\begin{eqnarray}
\label{pd1}
&
\frac{\phi_{i+1,j}^{n,s}(1+0.5/i)-2\phi_{i,j}^{n,s+1}+\phi_{i-1,j}^{n,s}(1-0.5/i)}{\Delta
    R^2}+\nonumber \\
&\frac{\phi_{i,j+1}^{n,s}-2\phi_{i,j}^{n,s+1}+\phi_{i,j-1}^{n,s}}{\Delta Z^2}
=-4\pi\rho_{i,j}^{n} \text{ for } i\neq 0,\nonumber \\
&
\frac{8\phi_{i+1,j}^{n,s}-8\phi_{i,j}^{n,s+1}}{\Delta
    R^2}+\\
&\frac{\phi_{i,j+1}^{n,s}-2\phi_{i,j}^{n,s+1}+\phi_{i,j-1}^{n,s}}{\Delta Z^2}
=-4\pi\rho_{i,j}^{n} \text{ for } i=0,\nonumber \\
&\phi_{i,j}^{n,0}=\phi_{i,j}^{n-1},\nonumber 
\end{eqnarray}
proven to be effective in plasmadynamics and for solving the
vacuum electronic problems \cite{Anishchenko2014}. Here,  $s$  is the
iteration number and $n$ is the time step number. Iterations in
\ref{pd3} occur until all the quantities
$|\phi^{n,s}-\phi^{n,s-1}|$ become less than $\epsilon U$. The
parameter $\epsilon\ll1$ is a small fraction of accelerating
voltage $U$. (In our simulations, $\epsilon$ is set to be equal to
$10^{-5}$.)

As a rule, iterative methods are slowly
converging~\cite{Hockney1981}. However, this is not always the
case in dynamical problems~\cite{Anishchenko2014}. The main reason is
proper choice of initial approximation in the iteration scheme.
Namely, the potential distribution at the previous time step $n-1$
plays the role of initial approximation at the step $n$. As a result, the Jacobi method reduces to
one-three iterations at each time step. The procedure requires
much less simulation time than, say, the computation of new
velocities and positions of the particles.

To solve the Poisson equation, we use the Dirichlet boundary
conditions. The conditions imply specification of potentials at
the cathode ($\phi=U_c$) and the anode ($\phi=U_a$) and at the
edge of the computational region ($z=0$). For the latter, the
logarithmic distribution is used  \cite{Bugaev1979}
\begin{equation}
\label{pd2}
\phi=U_c+\frac{(U_a-U_c)\ln(r/r_c)}{\ln(r_a/r_c)}.
\end{equation}
This distribution exactly describes $\phi$ change in the gap
between two infinite cylinders. The potential distribution
approaches \ref{pd5} if the edge of the computational region
($z=0$) is located at a considerable distance from the
electron-emitting surface.

Charge density $\rho_{i,j}^{n}$ can be found with the linear
weighting procedure \cite{Birdsall1985}. This means that each particle contributes to
the charge density of the four nearest nodes according to the
following formulas
\begin{eqnarray}
\label{pd3}
&\rho_{i,j}^n=\sum_\alpha\Delta\rho_{i,j}^{\alpha,n},\nonumber \\
&\Delta\rho_{i,j}^{\alpha n}=\frac{q_\alpha}{\Delta V}\Big(1-\frac{(r_\alpha^n)^2-(\Delta Ri)^2}{\Delta R^2(2i+1)}\Big)\Big(1-\frac{z_\alpha^n-\Delta Zj}{\Delta
    Z}\Big),\nonumber \\
&\Delta\rho_{i+1,j}^{\alpha n}=\frac{q_\alpha}{\Delta V}\frac{(r_\alpha^n)^2-(\Delta Ri)^2}{\Delta R^2(2i+1)}\Big(1-\frac{z_\alpha^n-\Delta Zj}{\Delta
    Z}\Big), \\
&\Delta\rho_{i,j+1}^{\alpha n}=\frac{q_\alpha}{\Delta V}\Big(1-\frac{(r_\alpha^n)^2-(\Delta Ri)^2}{\Delta R^2(2i+1)}\Big)\frac{z_\alpha^n-\Delta Zj}{\Delta
    Z},\nonumber \\
&\Delta\rho_{i+1,j+1}^{\alpha n}=\frac{q_\alpha}{\Delta V}\frac{(r_\alpha^n)^2-(\Delta Ri)^2}{\Delta R^2(2i+1)}\frac{z_\alpha^n-\Delta Zj}{\Delta
    Z}.\nonumber 
\end{eqnarray}
Here,
\begin{eqnarray}
&\Delta V=\pi \Delta R^2\Delta Z(2j+1) \textbf{ for } i\neq 0, \nonumber \\
&\Delta V=\frac{\pi \Delta R^2\Delta Z}{4}
\textbf{ for } i=0, \nonumber
\end{eqnarray}
$q_\alpha$ and $\vec r_\alpha^n=(r_\alpha^n,z_\alpha^n)$ are particles' charges and positions, respectively.

The electric field is related to electrostatic
potential~$\phi_{i,j}^n$ as
\begin{equation}
\vec E^{n}_{i,j}=-\bigg(\frac{\phi^n_{i+1,j}-\phi^n_{i-1,j}}{2\Delta
    R},\frac{\phi^n_{i,j+1}-\phi^n_{i,j-1}}{2\Delta Z}\bigg).
\end{equation}

A non-rotating axially symmetric electron beam possesses only the
polar component of the self-induced magnetic field $H$. The
component $H$ can be found using the Stokes theorem:
\begin{equation}
H=\frac{4\pi\int_0^rj_z(r_1,z)r_1dr_1}{cr}.
\end{equation}
(To calculate the magnetic field properly, one must include the current in the cathode itself~\cite{Poukey1973,Bugaev1979}.)

Note that the axial component of the current density can be found
by the same weighting procedure as the charge density. It is only
necessary to substitute $q_\alpha v_{z\alpha}^n$ instead of
$q_\alpha$ in \ref{pd3}.

Numerical integration of the Newton-Lorentz equations is the most
time-consuming procedure. That is why the integration algorithm
must be as fast as possible. During several decades the Boris
algorithm was used for this purpose. However, this algorithm can
lead to spurious forces acting on particles~\cite{Vay2008}.  To overcome the
drawback, a new algorithm for momenta integration has been
proposed by J.-L. Vay. In the axially symmetric
case, the numerical scheme can be written as follows
\begin{eqnarray}
\label{pd4}
& p_{r\alpha}^{n+1/2}=p_{r\alpha}^n+q_\alpha\Big(E_{r\alpha}^{n+1/2}-\frac{v_{z\alpha}^n}{2c}H_\alpha^{n+1/2}\Big)\Delta T,\nonumber \\
& p_{z\alpha}^{n+1/2}=p_{z\alpha}^n+q_\alpha\Big(E_{z\alpha}^{n+1/2}+\frac{v_{r\alpha}^n}{2c}H_\alpha^{n+1/2}\Big)\Delta T,\nonumber \\
&\gamma_\alpha^{n+1}=\sqrt{1+\frac{(p_{r\alpha}^n+q_\alpha E_{r\alpha}^{n+1/2}\Delta T)^2}{m_\alpha^2c^2}+\frac{(p_{z\alpha}^n+q_\alpha E_{z\alpha}^{n+1/2}\Delta T)^2}{m_\alpha^2c^2}},\nonumber \\
& A_\alpha=\frac{q_\alpha H_\alpha^n\Delta T}{2m_\alpha c\gamma_\alpha^{n+1}},\nonumber \\
& B_\alpha=1+A_\alpha^2, \\
& p_{r\alpha}^{n+1}=(p_{r\alpha}^{n+1/2}-A_\alpha p_{z\alpha}^{n+1/2})/B_\alpha,\nonumber \\
& p_{z\alpha}^{n+1}=(p_{z\alpha}^{n+1/2}+A_\alpha p_{r\alpha}^{n+1/2})/B_\alpha,\nonumber \\
& v_{r\alpha}^{n+1}=\frac{p_{r\alpha}^{n+1}}{\gamma_\alpha^{n+1}m_\alpha},\nonumber \\
& v_{z\alpha}^{n+1}=\frac{p_{z\alpha}^{n+1}}{\gamma_\alpha^{n+1}m_\alpha}.\nonumber
\end{eqnarray}
Here, $E_{r\alpha}^{n+1/2}$, $E_{z\alpha}^{n+1/2}$, and
$H_{\alpha}^{n+1/2}$ are the electric and magnetic fields acting
on a particle with momenta $\vec p_\alpha$ and velocity $\vec
v_\alpha$ at time $\Delta T(n+1/2)$.

 In quasi-stationary flows, the fields are slowly varying functions of time.
This property gives us the possibility to use $E_{r\alpha}^{n}$,
$E_{z\alpha}^{n}$,  and $E_{z\alpha}^{n}$ instead of
$E_{r\alpha}^{n+1/2}$, $E_{z\alpha}^{n+1/2}$, and
$H_{\alpha}^{n+1/2}$, respectively \cite{Poukey1973}. The former
quantities can be found by weighting procedure \cite{Birdsall1985}, for example,
\begin{eqnarray}
\label{pd5}
& H_{\alpha}^{n}=H_{i,j}\Big(1-\frac{(r_\alpha^n)^2-(\Delta Ri)^2}{\Delta R^2(2i+1)}\Big)\Big(1-\frac{z_\alpha^n-\Delta Zj}{\Delta
    Z}\Big),\nonumber \\
&+H_{i+1,j}\frac{(r_\alpha^n)^2-(\Delta Ri)^2}{\Delta R^2(2i+1)}\Big(1-\frac{z_\alpha^n-\Delta Zj}{\Delta
    Z}\Big), \\
&+H_{i,j+1}\Big(1-\frac{(r_\alpha^n)^2-(\Delta Ri)^2}{\Delta R^2(2i+1)}\Big)\frac{z_\alpha^n-\Delta Zj}{\Delta
    Z},\nonumber \\
&+H_{i+1,j+1}\frac{(r_\alpha^n)^2-(\Delta Ri)^2}{\Delta R^2(2i+1)}\frac{z_\alpha^n-\Delta Zj}{\Delta
    Z},\nonumber
\end{eqnarray}
where magnetic field $H$ is calculated in four nearest to the
particle nodes.

Using central difference approximations for velocities,
\begin{eqnarray}
& v_{r\alpha}^{n+1/2}=\frac{1}{2}(v_{r\alpha}^n+v_{r\alpha}^{n+1}),\nonumber \\
& v_{z\alpha}^{n+1/2}=\frac{1}{2}(v_{z\alpha}^n+v_{z\alpha}^{n+1}),\nonumber
\end{eqnarray}
we update the positions \cite{Poukey1973},
\begin{eqnarray}
& r_\alpha^{n+1}=r_\alpha^n+v_{r\alpha}^{n+1/2}\Delta T,\nonumber \\
& z_\alpha^{n+1}=r_\alpha^n+v_{z\alpha}^{n+1/2}\Delta T.\nonumber
\end{eqnarray}

Two types of particle injection are  implemented in the simulation
code, namely, overinjection and the dual cell algorithm described
in \cite{Watrous2001}. They both lead to the similar results.

In simulations, a simple uniform grid with $\Delta R=\Delta
Z=100$~$\mu$m  is used. The time step $\Delta T$ is set to be
equal to $0.5\Delta Z/c\approx167$~fs.

\bibliography{references}

\begin{thebibliography}{18}%
\makeatletter
\providecommand \@ifxundefined [1]{%
 \@ifx{#1\undefined}
}%
\providecommand \@ifnum [1]{%
 \ifnum #1\expandafter \@firstoftwo
 \else \expandafter \@secondoftwo
 \fi
}%
\providecommand \@ifx [1]{%
 \ifx #1\expandafter \@firstoftwo
 \else \expandafter \@secondoftwo
 \fi
}%
\providecommand \natexlab [1]{#1}%
\providecommand \enquote  [1]{``#1''}%
\providecommand \bibnamefont  [1]{#1}%
\providecommand \bibfnamefont [1]{#1}%
\providecommand \citenamefont [1]{#1}%
\providecommand \href@noop [0]{\@secondoftwo}%
\providecommand \href [0]{\begingroup \@sanitize@url \@href}%
\providecommand \@href[1]{\@@startlink{#1}\@@href}%
\providecommand \@@href[1]{\endgroup#1\@@endlink}%
\providecommand \@sanitize@url [0]{\catcode `\\12\catcode `\$12\catcode
  `\&12\catcode `\#12\catcode `\^12\catcode `\_12\catcode `\%12\relax}%
\providecommand \@@startlink[1]{}%
\providecommand \@@endlink[0]{}%
\providecommand \url  [0]{\begingroup\@sanitize@url \@url }%
\providecommand \@url [1]{\endgroup\@href {#1}{\urlprefix }}%
\providecommand \urlprefix  [0]{URL }%
\providecommand \Eprint [0]{\href }%
\providecommand \doibase [0]{http://dx.doi.org/}%
\providecommand \selectlanguage [0]{\@gobble}%
\providecommand \bibinfo  [0]{\@secondoftwo}%
\providecommand \bibfield  [0]{\@secondoftwo}%
\providecommand \translation [1]{[#1]}%
\providecommand \BibitemOpen [0]{}%
\providecommand \bibitemStop [0]{}%
\providecommand \bibitemNoStop [0]{.\EOS\space}%
\providecommand \EOS [0]{\spacefactor3000\relax}%
\providecommand \BibitemShut  [1]{\csname bibitem#1\endcsname}%
\let\auto@bib@innerbib\@empty
\bibitem [{\citenamefont {Booske}(2011)}]{Booske2011}%
  \BibitemOpen
  \bibfield  {author} {\bibinfo {author} {\bibfnamefont {J.~H.}\ \bibnamefont
  {Booske}},\ }\href@noop {} {\bibfield  {journal} {\bibinfo  {journal} {IEEE
  Trans. Terahertz Sci. Technol.}\ }\textbf {\bibinfo {volume} {1}},\ \bibinfo
  {pages} {54} (\bibinfo {year} {2011})}\BibitemShut {NoStop}%
\bibitem [{\citenamefont {Dhillon}\ and\ \citenamefont
  {et~al.}(2017)}]{Dhillon2017}%
  \BibitemOpen
  \bibfield  {author} {\bibinfo {author} {\bibfnamefont {S.~S.}\ \bibnamefont
  {Dhillon}}\ and\ \bibinfo {author} {\bibnamefont {et~al.}},\ }\href@noop {}
  {\bibfield  {journal} {\bibinfo  {journal} {J. Phys. D: Appl. Phys.}\
  }\textbf {\bibinfo {volume} {50}},\ \bibinfo {pages} {043001} (\bibinfo
  {year} {2017})}\BibitemShut {NoStop}%
\bibitem [{\citenamefont {Garate}\ and\ \citenamefont
  {et~al.}(1987)}]{Garate1987}%
  \BibitemOpen
  \bibfield  {author} {\bibinfo {author} {\bibfnamefont {E.~P.}\ \bibnamefont
  {Garate}}\ and\ \bibinfo {author} {\bibnamefont {et~al.}},\ }\href@noop {}
  {\bibfield  {journal} {\bibinfo  {journal} {Nucl. Instrum. Methods A}\
  }\textbf {\bibinfo {volume} {259}},\ \bibinfo {pages} {125} (\bibinfo {year}
  {1987})}\BibitemShut {NoStop}%
\bibitem [{\citenamefont {Bennett}(1934)}]{Bennet1934}%
  \BibitemOpen
  \bibfield  {author} {\bibinfo {author} {\bibfnamefont {W.~H.}\ \bibnamefont
  {Bennett}},\ }\href@noop {} {\bibfield  {journal} {\bibinfo  {journal} {Phys.
  Rev.}\ }\textbf {\bibinfo {volume} {45}},\ \bibinfo {pages} {890} (\bibinfo
  {year} {1934})}\BibitemShut {NoStop}%
\bibitem [{\citenamefont {Morrov}(1971)}]{Morrov1971}%
  \BibitemOpen
  \bibfield  {author} {\bibinfo {author} {\bibfnamefont {D.~L.}\ \bibnamefont
  {Morrov}},\ }\href@noop {} {\bibfield  {journal} {\bibinfo  {journal} {J.
  Appl. Phys.}\ }\textbf {\bibinfo {volume} {19}},\ \bibinfo {pages} {441}
  (\bibinfo {year} {1971})}\BibitemShut {NoStop}%
\bibitem [{\citenamefont {Bradley}\ and\ \citenamefont
  {Kuswa}(1972)}]{Bradley1972}%
  \BibitemOpen
  \bibfield  {author} {\bibinfo {author} {\bibfnamefont {L.~P.}\ \bibnamefont
  {Bradley}}\ and\ \bibinfo {author} {\bibfnamefont {G.~W.}\ \bibnamefont
  {Kuswa}},\ }\href@noop {} {\bibfield  {journal} {\bibinfo  {journal} {Phys.
  Rev. Lett.}\ }\textbf {\bibinfo {volume} {29}},\ \bibinfo {pages} {1441}
  (\bibinfo {year} {1972})}\BibitemShut {NoStop}%
\bibitem [{\citenamefont {Anishchenko}\ \emph {et~al.}(2017)\citenamefont
  {Anishchenko}, \citenamefont {Baryshevsky}, \citenamefont {Belous},
  \citenamefont {Gurinovich}, \citenamefont {Gurinovich}, \citenamefont
  {Gurnevich},\ and\ \citenamefont {Molchanov}}]{Anishchenko2017}%
  \BibitemOpen
  \bibfield  {author} {\bibinfo {author} {\bibfnamefont {S.}~\bibnamefont
  {Anishchenko}}, \bibinfo {author} {\bibfnamefont {V.}~\bibnamefont
  {Baryshevsky}}, \bibinfo {author} {\bibfnamefont {N.}~\bibnamefont {Belous}},
  \bibinfo {author} {\bibfnamefont {A.}~\bibnamefont {Gurinovich}}, \bibinfo
  {author} {\bibfnamefont {E.}~\bibnamefont {Gurinovich}}, \bibinfo {author}
  {\bibfnamefont {E.}~\bibnamefont {Gurnevich}}, \ and\ \bibinfo {author}
  {\bibfnamefont {P.}~\bibnamefont {Molchanov}},\ }\href@noop {} {\bibfield
  {journal} {\bibinfo  {journal} {IEEE Trans. Plasma Sci.}\ }\textbf {\bibinfo
  {volume} {45}},\ \bibinfo {pages} {2739} (\bibinfo {year}
  {2017})}\BibitemShut {NoStop}%
\bibitem [{\citenamefont {Anishchenko}\ and\ \citenamefont
  {Gurinovich}(2014)}]{Anishchenko2014}%
  \BibitemOpen
  \bibfield  {author} {\bibinfo {author} {\bibfnamefont {S.}~\bibnamefont
  {Anishchenko}}\ and\ \bibinfo {author} {\bibfnamefont {A.}~\bibnamefont
  {Gurinovich}},\ }in\ \href@noop {} {\emph {\bibinfo {booktitle} {Proc. 5th
  Euro-Asian Pulsed-Power Conf.}}}\ (\bibinfo {year} {2014})\ pp.\ \bibinfo
  {pages} {1--6}\BibitemShut {NoStop}%
\bibitem [{\citenamefont {Bugaev}\ and\ \citenamefont
  {et~al}(1979)}]{Bugaev1979}%
  \BibitemOpen
  \bibfield  {author} {\bibinfo {author} {\bibfnamefont {S.~P.}\ \bibnamefont
  {Bugaev}}\ and\ \bibinfo {author} {\bibnamefont {et~al}},\ }in\ \href@noop {}
  {\emph {\bibinfo {booktitle} {High-frequency relativistic electronics}}}\
  (\bibinfo {year} {1979})\ pp.\ \bibinfo {pages} {5--75}\BibitemShut {NoStop}%
\bibitem [{\citenamefont {Poukey}\ \emph {et~al.}(1973)\citenamefont {Poukey},
  \citenamefont {Freeman},\ and\ \citenamefont {Yonas}}]{Poukey1973}%
  \BibitemOpen
  \bibfield  {author} {\bibinfo {author} {\bibfnamefont {J.~W.}\ \bibnamefont
  {Poukey}}, \bibinfo {author} {\bibfnamefont {J.~R.}\ \bibnamefont {Freeman}},
  \ and\ \bibinfo {author} {\bibfnamefont {G.}~\bibnamefont {Yonas}},\
  }\href@noop {} {\bibfield  {journal} {\bibinfo  {journal} {J. Vac. Sci.
  Technol.}\ }\textbf {\bibinfo {volume} {10}},\ \bibinfo {pages} {954}
  (\bibinfo {year} {1973})}\BibitemShut {NoStop}%
\bibitem [{\citenamefont {Golovin}(1989)}]{Golovin1989}%
  \BibitemOpen
  \bibfield  {author} {\bibinfo {author} {\bibfnamefont {G.~T.}\ \bibnamefont
  {Golovin}},\ }\href@noop {} {\bibfield  {journal} {\bibinfo  {journal}
  {U.S.S.R. Comput. Math. Math. Phys.}\ }\textbf {\bibinfo {volume} {29}},\
  \bibinfo {pages} {67} (\bibinfo {year} {1989})}\BibitemShut {NoStop}%
\bibitem [{\citenamefont {Roshal}(1979)}]{Roshal1979}%
  \BibitemOpen
  \bibfield  {author} {\bibinfo {author} {\bibfnamefont {A.~S.}\ \bibnamefont
  {Roshal}},\ }\href@noop {} {\emph {\bibinfo {title} {Modeling of charged
  beams}}}\ (\bibinfo  {publisher} {Atomizdat, Moscow},\ \bibinfo {year}
  {1979})\ \bibinfo {note} {[in Russian]}\BibitemShut {NoStop}%
\bibitem [{\citenamefont {Hockney}\ and\ \citenamefont
  {Easwood}(1981)}]{Hockney1981}%
  \BibitemOpen
  \bibfield  {author} {\bibinfo {author} {\bibfnamefont {R.~W.}\ \bibnamefont
  {Hockney}}\ and\ \bibinfo {author} {\bibfnamefont {J.~W.}\ \bibnamefont
  {Easwood}},\ }\href@noop {} {\emph {\bibinfo {title} {Computer simulation
  using particles}}}\ (\bibinfo  {publisher} {McGraw-Hill, New York},\ \bibinfo
  {year} {1981})\BibitemShut {NoStop}%
\bibitem [{\citenamefont {Birdsall}\ and\ \citenamefont
  {Langdon}(1985)}]{Birdsall1985}%
  \BibitemOpen
  \bibfield  {author} {\bibinfo {author} {\bibfnamefont {C.~K.}\ \bibnamefont
  {Birdsall}}\ and\ \bibinfo {author} {\bibfnamefont {A.~B.}\ \bibnamefont
  {Langdon}},\ }\href@noop {} {\emph {\bibinfo {title} {Plasma physics, via
  computer simulations}}}\ (\bibinfo  {publisher} {McGraw-Hill, New York},\
  \bibinfo {year} {1985})\BibitemShut {NoStop}%
\bibitem [{\citenamefont {Antonsen}\ and\ \citenamefont
  {et~al.}(1999)}]{Antonsen}%
  \BibitemOpen
  \bibfield  {author} {\bibinfo {author} {\bibfnamefont {T.~M.}\ \bibnamefont
  {Antonsen}}\ and\ \bibinfo {author} {\bibnamefont {et~al.}},\ }\href@noop {}
  {\bibfield  {journal} {\bibinfo  {journal} {Proceedings of the IEEE.}\
  }\textbf {\bibinfo {volume} {87}},\ \bibinfo {pages} {804} (\bibinfo {year}
  {1999})}\BibitemShut {NoStop}%
\bibitem [{\citenamefont {Verboncoeur}(2005)}]{Verboncoeur2005}%
  \BibitemOpen
  \bibfield  {author} {\bibinfo {author} {\bibfnamefont {J.~P.}\ \bibnamefont
  {Verboncoeur}},\ }\href@noop {} {\bibfield  {journal} {\bibinfo  {journal}
  {Plasma Phys. Control. Fusion.}\ }\textbf {\bibinfo {volume} {47}},\ \bibinfo
  {pages} {231} (\bibinfo {year} {2005})}\BibitemShut {NoStop}%
\bibitem [{\citenamefont {Vay}(2008)}]{Vay2008}%
  \BibitemOpen
  \bibfield  {author} {\bibinfo {author} {\bibfnamefont {J.-L.}\ \bibnamefont
  {Vay}},\ }\href@noop {} {\bibfield  {journal} {\bibinfo  {journal} {Phys.
  Plasmas}\ }\textbf {\bibinfo {volume} {15}},\ \bibinfo {pages} {056701}
  (\bibinfo {year} {2008})}\BibitemShut {NoStop}%
\bibitem [{\citenamefont {Watrous}\ \emph {et~al.}(2001)\citenamefont
  {Watrous}, \citenamefont {Lugisland},\ and\ \citenamefont
  {Sasser}}]{Watrous2001}%
  \BibitemOpen
  \bibfield  {author} {\bibinfo {author} {\bibfnamefont {J.~J.}\ \bibnamefont
  {Watrous}}, \bibinfo {author} {\bibfnamefont {J.~W.}\ \bibnamefont
  {Lugisland}}, \ and\ \bibinfo {author} {\bibfnamefont {G.~E.}\ \bibnamefont
  {Sasser}},\ }\href@noop {} {\bibfield  {journal} {\bibinfo  {journal} {Phys.
  Plasmas}\ }\textbf {\bibinfo {volume} {8}},\ \bibinfo {pages} {289} (\bibinfo
  {year} {2001})}\BibitemShut {NoStop}%
\end{thebibliography}%

\end{document}